\documentclass[twocolumn,amsmath,amssymb,floatfix,showpacs,superscriptaddress,nofootinbib,longbibliography]{revtex4-2}
\usepackage[dvipsnames]{xcolor}
\usepackage[colorlinks=true,linktoc=page,linkcolor=blue,citecolor=blue,urlcolor=blue]{hyperref}
\usepackage{float,subfigure,physics,array,amsmath,bigints,orcidlink,marvosym,comment}
\newcommand{\la}{{\langle}}
\newcommand{\ra}{{\rangle}}
\allowdisplaybreaks

\begin{document}
	
\title{Quantum Otto heat-engine with Kitaev-Heisenberg cluster: Possible roles of frustration, magnons, and duality}

\author{Sheikh Moonsun Pervez~\orcidlink{0009-0000-2283-053X}}
\email[corresponding author,~\Letter~]{moonsun@iopb.res.in}
\affiliation{Institute of Physics, Sachivalaya Marg, Bhubaneswar-751005, India}
\affiliation{Homi Bhabha National Institute, Training School Complex, Anushakti Nagar, Mumbai 400094, India}

\author{Saptarshi Mandal~\orcidlink{0000-0001-8261-0483}}
\email[\Letter~]{saptarshi@iopb.res.in}
\affiliation{Institute of Physics, Sachivalaya Marg, Bhubaneswar-751005, India}
\affiliation{Homi Bhabha National Institute, Training School Complex, Anushakti Nagar, Mumbai 400094, India}


\begin{abstract}
We study the performance of Kitaev-Heisenberg (KH) clusters as working media realizing a quantum Otto engine (QOE). An external Zeeman field with linear time dependency is used as the driving mechanism. The efficiency strongly depends on Kitaev ($\kappa$) and Heisenberg ($J$) exchange interaction. Interestingly, efficiency is comparable when the relative magnitude of $\kappa$ and $J$ is the same but of opposite signs. The above results are explained due to a subtle interplay of frustration, quantum fluctuation, and duality of eigen-spectra for the KH system when both the signs of $\kappa$ and $J$ are reversed. The maximum efficiency is shown to be dynamically related to eigen-spectra forming discrete narrow bands, where total spin angular momentum becomes a good quantum number. We relate this optimum efficiency to the realization of weakly interacting magnons, where the system reduces to an approximate eigen-system of the external drive. Finally, we extend our study to the large spin Kitaev model and find a quantum advantage in efficiency for $S=1/2$. The results could be of practical interest for materials with KH interactions as a platform for QOE. 
\end{abstract}

\date{\today}
\maketitle


\section{Introduction}\label{section_introduction}

The rapid advancement of technology has made it possible to realize interesting quantum devices, including nano-sensors~\cite{quantum_sensing} and nano-magnetism~\cite{RevModPhys.81.1495}, where the quantum effect is the key controlling mechanism. In this respect, recent studies on quantum heat engine (QHE)~\cite{quantum_osxillator_HE,continuous_devices,fidelity_definition, PhysRevA.107.L040202,robust_quantum_refrigerator,quasi_static_limits,kibble_zurek_scaling} and batteries~\cite{lindblad_equation,multimode_advantage,dimensional_enhancement, QB_non_hermitian_charging,fast_charging} have revealed many aspects of thermodynamics~\cite{Binder2018,Jarzynski-1997,campisi-RevModPhys.83.771,Aberg-PhysRevX.8.011019,fernando-PhysRevLett.111.250404,efficiency_beyond_second_law} in small systems, where various quantum effects compete with the general rules of thermodynamics, and often results in counter intuitive outcome which is absent in macroscopic, classical heat engines. Compared to conventional heat engines, the working medium (WM) of QHE promises to be more versatile because of the possibility of greater control to manipulate quantum states of matter ~\cite{dalessandro2008introduction,nielsen2010quantum,konrad-PhysRevA.104.052614,Julian-PRXQuantum.2.030335,roberto-sym15112033,Gour2018}. Along with the recent advancement of nano-science and low-temperature physics, the effect of quantum thermodynamics is inevitable, and it renders the study of QHE an important field of research in present days. Starting from the original proposal of QHE in the context of maser~\cite{scovil-prl-1959,scovil-pr-1967}, recent theoretical and experimental studies show remarkable advancement in understanding quantum thermodynamics at the small scale~\cite{Horodecki2013,Skrzypczyk2014,doi:10.1073/pnas.1411728112,Lostaglio2015,muller-PhysRevX.8.041051,PhysRevX.8.021064}.
\begin{figure}[h]\centering
	\includegraphics[width=\columnwidth,height=!]{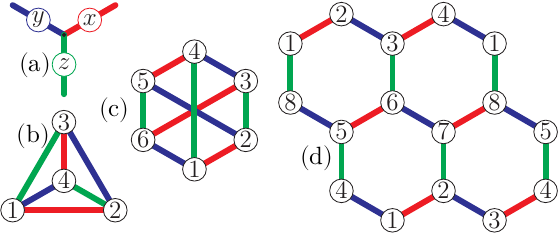}
	\caption{(a) The red, blue, and green bonds indicate $x$-$x$, $y$-$y$, $z$-$z$-type interactions in Kitaev model. Panels (b), (c), and (d) depict the four, six, and eight-site clusters considered, respectively.}\label{figure_schematics_clusters}
\end{figure}
\\\\\indent
In this work, we study the quantum Otto engine (QOE)~\cite{arnaud2003simplequantumheatengine,Tien-2004-PRL.93.140403} taking small Kitaev-Heisenberg (KH) clusters~\cite{PhysRevB.101.115133, Pervez2025, MoonsunPervez2025,doi:10.7566/JPSJ.89.114709,wilson2025exacteigenvaluesexperimentalsignatures} as WM, and follow the Otto cycle~\cite{FNori-PhysRevE.76.031105}. Otto engines deliver better efficiency with less emission, and have potential applications ranging from small generators to large-scale industrial applications~\cite{heywood,peterson,stone,revisiting_otto}. Initial works on QOE mainly considered non-interacting quantum systems such as harmonic oscillator, two or few level atomic or spin systems, which have the convenience of analytical tractability~\cite{Feldman-PhysRevE.68.016101,Quan-PhysRevE.72.056110,Nori-PhysRevLett.97.180402,Rezek-2006,Feldmann-1996,Eitan-1992,e19040136}. The study on ideal quantum gas shows that, while ideal Bose gas exceeds the work and efficiency than that of classical ideal gas, ideal Fermi gas performs below classical limit~\cite{sisman-2001,Chen-PhysRevE.98.062119,WANG2009706,wang-Jap-2009,PhysRevE.79.041129}. However, these simple quantum systems mostly rely on single-particle quantum properties. To explore the true many-particle aspects in QOE performance, various studies have been recently carried out, such as one-dimensional Heisenberg-chain~\cite{johal-symmetry-2021}, Ising system~\cite{open_ising_chain_QE,ising_critical}, XYZ-chain with DM interactions~\cite{XY_CHAIN_WITH_dm,Kuznetsova2023}, and one-dimensional Kitaev chains~\cite{long_range_interaction,long_range_advantage}. All these studies reveal many interesting many-particle quantum aspects such as geometric friction~\cite{geometric_friction}, topological and finite-size effects ~\cite{topologicalfinitesizeeffects}, etc. Experiments on the realization of QOE~\cite{effective_negative_temperature,experimental_spin_quantum_engine,single_qubit} have also been achieved successfully, elucidating the concept of effective negative temperatures ~\cite{effective_negative_temperature,enhancedefficiencyquantumotto}.
\\\\\indent
The WM of this study, viz., the KH system, is one of the paradigmatic models exhibiting unique quantum state of matter, known as the quantum spin liquid (QSL)~\cite{review_balents_nature,review_balents,mandaljpa,kitaev_2006,Mandal_2025,naveen_2008,PhysRevB.90.104424,PhysRevB.102.134309,eschmann_2020}. Efforts of material realization of the Kitaev-QSL have progressed in positive direction in various Mott insulators~\cite{jackeli_khaliullin, review_knolle,review_lee,review_trebst,takagi_2019,motome_2020,hermanns_2018,abanerjee_2016}, such as $\alpha$-RuCl$_3$, Na$_2$IrO$_3$, Li$_2$IrO$_3$, etc.~\cite{banerjee_2017_science,PhysRevLett.110.076402,PhysRevLett.108.127204,PhysRevLett.114.077202,PhysRevLett.108.127203}. Here we discover interesting many-body effects of frustrated spin-system controlling the efficiency and work of the QOE for KH system. Our plan for the presentation is as follows.
\\\\\indent
In \ref{subsection_hamiltonian}, we introduce the WM and briefly describe possible regimes of the system that are relevant to describing the performances of QOE. The four-stroke Otto-cycle is introduced in \ref{subsection_otto_cycle}. The mathematical definition of efficiency and all possible operational modes that a QOE can perform are also mentioned. In \ref{section_results}, we chart out the key results in various subsections. In \ref{subsection_working_modes}, we introduce the relevant parameter spaces for the Kitaev clusters for all four different operational modes (engine, accelerator, heater, refrigerator), and characterize them in terms of some fictitious effective temperature. In \ref{subsection_heisenberg_interaction}, we describe the behavior of maximum efficiency achieved and how it depends on the external magnetic field and relative signs of Kitaev and Heisenberg interactions. We successfully relate this maximum efficiency and other key properties to some interesting many-body aspects, such as the bunching effect of eigen-spectrum and duality property of high energy and low energy spectrum of KH clusters when the sign of both these interactions is reversed. In \ref{subsection_large_s}, we show large spin-$S$ Kitaev cluster performance as QOE and find that there is a quantum advantage, i.e., the efficiency decreases as $S$ is increased. The same quantum advantage is found to be true even for pure Heisenberg interaction in a tri-coordinated cluster. We conclude in section \ref{section_conclusion}.


\section{Model and method}\label{section_model_and_method}
\subsection{Hamiltonian}\label{subsection_hamiltonian}

{\it Kitaev clusters}: The WM under consideration are KH clusters of size 4, 6, and 8-sites, as schematically shown in FIG.~\ref{figure_schematics_clusters}. Hamiltonian of the system can be written as~\cite{PhysRevB.99.140413,PhysRevB.97.241110,PhysRevB.98.060416,PhysRevB.101.180401,PhysRevB.108.035149,PhysRevB.108.165118,lyu2025multipartyentanglementloopsquantum},
\begin{eqnarray}\label{hamiltonian}
	&&H(t)=H_{\rm K}+ H_{\rm H} -h(t)\sum_{j}S_{j}^{z}, \\
	&&H_{\rm K}=\kappa \sum_{{\langle j, k \rangle}_{\alpha}} S_{j}^{\alpha} S_{k}^{\alpha},~H_{\rm H}=J\sum_{{\langle j, k \rangle}} \vec{S}_i \cdot \vec{S}_j ,~~ \alpha=x,y,z.~~~~
\end{eqnarray}
Here $\vec{S}_j$ denotes the spin 1/2 operator at site `$j$'. $H_{\rm K}$ and $H_{\rm H}$ are the (time-independent) Kitaev and Heisenberg interactions, respectively. The last term in Eq.(\ref{hamiltonian}) denotes the external driving Hamiltonian, and for simplicity, we consider a linear time dependency of $h(t)$. The index $\alpha$ in $H_{\rm K}$ denotes the $\alpha$-type bonds used to enumerate Kitaev interactions [FIG.~\ref{figure_schematics_clusters}]. $\kappa$ and $J$ denote the isotropic Kitaev and Heisenberg interactions. Positive $\kappa$ or $J$ indicates antiferromagnetic (AFM) interaction, and their negative values correspond to ferromagnetic (FM) coupling. In our study, all possible signs and relevant magnitudes of $\kappa$ and $J$ are considered. 
\begin{figure}[h]\centering
	\includegraphics[width=\columnwidth,height=!]{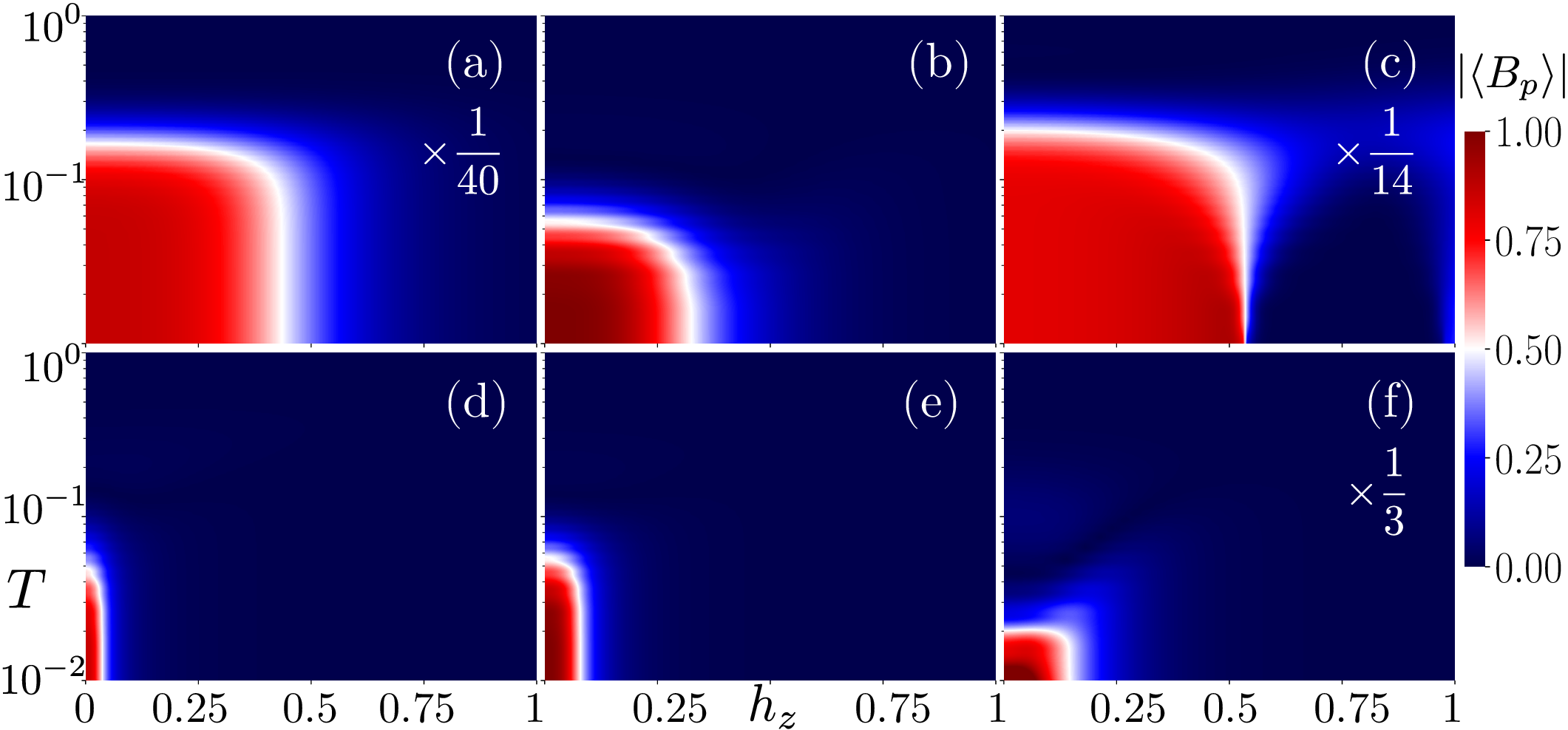}
	\caption{ $|\la B_p\ra|$ in magnetic field ($h_z$) - temperature ($T$) plane for 8-site cluster is plotted. The upper and lower panels are for the AFM and FM Kitaev model, respectively. Value of $J$ is taken to be -0.3, 0, and 0.3 in panels (a,d), (b,e), and (c,f), respectively. For better visualization, panels (a), (c), and (f) have been scaled up; the actual data correspond to the colored data multiplied by the factor indicated in the respective panels.}\label{figure_bp}
\end{figure}
\begin{figure}[h]\centering
	\includegraphics[width=\columnwidth,height=!]{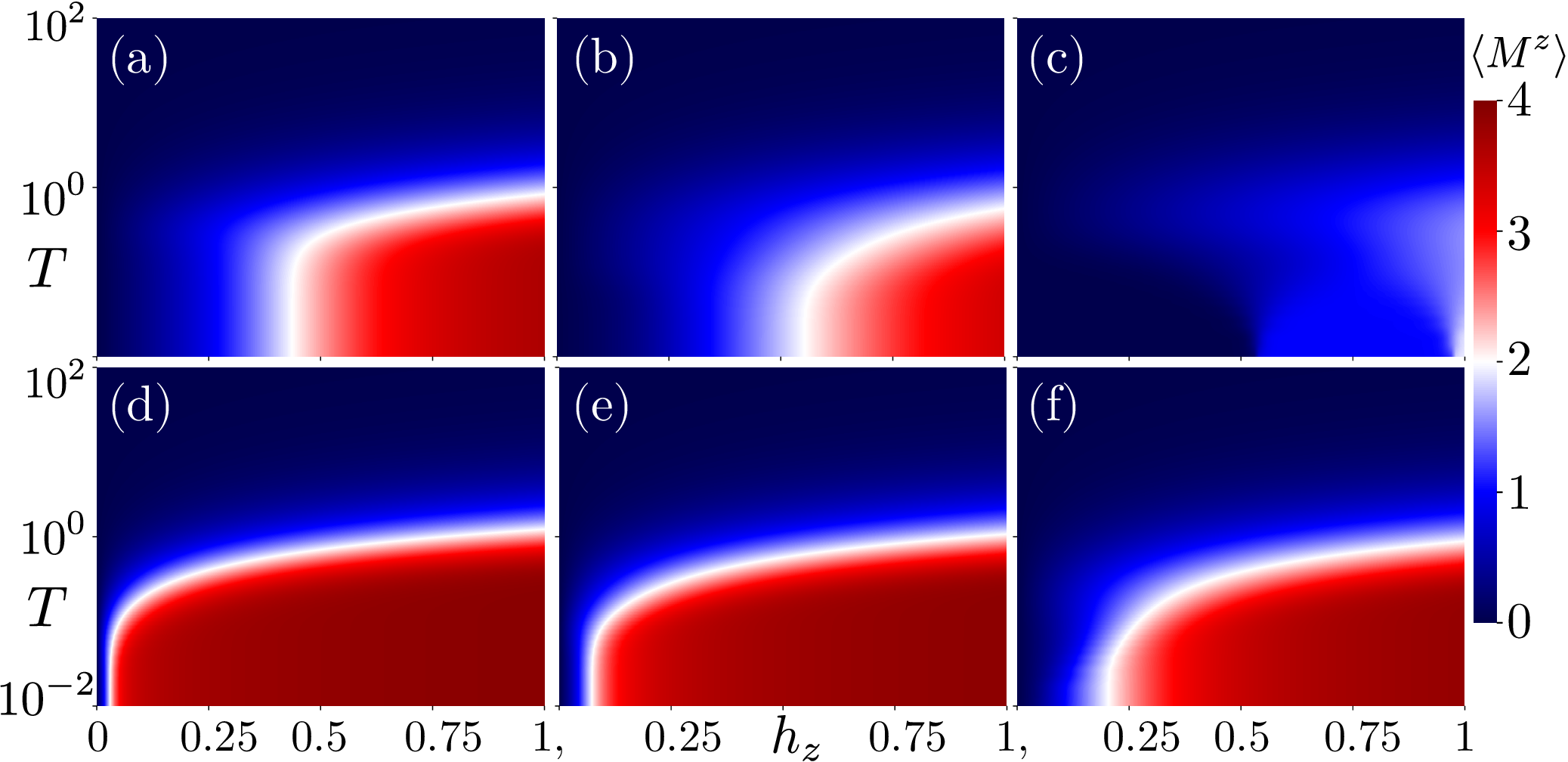}
	\caption{Total magnetization $\langle M^z \rangle$ for 8-site cluster is plotted in $h_z-T$ plane. The upper and lower panels correspond to $\kappa=1$ and $\kappa=-1$, respectively. Panels (a,d), (b,e), and (c,f) correspond to $J=-0.3, 0, 0.3$, respectively. }\label{figure_mz}
\end{figure}
\\\\\indent
{\it Brief description of possible regimes of Kitaev-Heisenberg clusters}: Here we briefly describe the possible regimes that occur at different values of external Zeeman field, for various signs and magnitudes of $\kappa$ and $J$. This will help us understand the dynamical aspects of various states of KH clusters that govern the QOE's performance. We consider two relevant order parameters that are complementary to each other at different limits. The pure Kitaev Hamiltonian $H_{\rm K}$ hosts a QSL state in the thermodynamic limit. This QSL is characterized by short-range correlation, vanishing magnetization, conserved flux operator $B_p$, and spin-fractionalization~\cite{kitaev_2006,saptarshi_2007}. $B_p$ is defined as the product of Kitaev Hamiltonian terms around a plaquette, and in our consideration, $B_p=\sigma_{1}^{z}\sigma_{2}^{y}\sigma_{7}^{x}\sigma_{6}^{z}\sigma_{5}^{y}\sigma_{4}^{x}$. On the other hand, the last two terms of Eq.(\ref{hamiltonian}) can be conveniently characterized by magnetization $\langle M^z \rangle$.
\\\\\indent
In FIG.~\ref{figure_bp}, comparison of panels (b) and (e) shows that $\langle B_p \rangle$ assumes a higher value (and close to 1), for a larger region for $\kappa=1$ than $\kappa=-1$, which is found for a Kitaev system even in the 2D limit~\cite{Hickey2019}. Addition of $J$ rapidly decreases $\langle B_p \rangle$ of the AFM model, as seen in (a) and (c). On the other hand, inclusion of $J$ for the FM model causes less decrement in comparison to the AFM one (panels (d) and (f)). A larger value of $\langle B_p \rangle$ corresponds to a proximate fractionalized phase with QSL state in the 2D limit. As we are dealing with clusters, we define this state as p-QSL, dubbed a `pro-quantum spin liquid state’. The p-QSL implies a possible realization of the QSL state in the thermodynamic limit. In FIG.~\ref{figure_mz}, we plot $\langle M^z \rangle$ in $T-h_z$ plane, for the parameters same as in FIG.~\ref{figure_bp}. There is a one-to-one correspondence of high $\langle B_p \rangle$ region to low $\la M^z\ra$ in the low temperature region. At high temperature, both quantities are vanishingly small, which denotes the usual paramagnetic phase in the 2D limit. As these small clusters qualitatively mimic the larger KH system very well, we predict the results of QOE shall remain similar for WM with more number of sites, which are difficult to access computationally, but may be relevant to potential applications in quantum technologies.

\subsection{Quantum Otto cycle}\label{subsection_otto_cycle}

\begin{figure}[h]\centering
	\includegraphics[width=0.7\columnwidth,height=!]{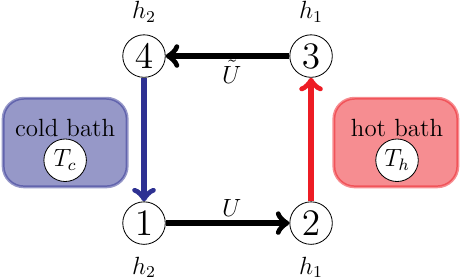}
	\caption{Otto cycle is shown schematically. At (1), the system is described by the initial magnetic field $h_2$ and density matrix $\rho^{(1)}$, in equilibrium with cold-bath. The system is then detached from the cold bath and unitarily evolved to (2) with a magnetic field linearly increased up to $h_1$, and characterized by a density matrix $\rho^{(2)}$. Now the system is brought in contact with a hot bath and described by a density matrix $\rho^{(3)}$. Finally, the system is detached from the hot bath and undergoes a unitary evolution to reach the state with $h_2$ with $\rho^{(4)}$ at (4).}
	\label{figure_otto_cycle}
\end{figure}

Here we briefly describe the principle of QOE, as shown in FIG.~\ref{figure_otto_cycle}. The Otto cycle involves a path $1$ $\rightarrow$ $2$ $\rightarrow$ $3$ $\rightarrow$ $4$ $\rightarrow$ $1$. The WM with initial Hamiltonian $H^{(1)}$ is made to equilibrate with a cold bath at temperature $T_c=1/\beta_c$, to have energy $E^{(1)}={\rm Tr}(\rho^{(1)} H^{(1)})$, where $\rho^{(1)}= e^{-\beta_c H^{(1)}}/{\rm Tr}(e^{-\beta_c H^{(1)}})$. Hamiltonian follows a unitary evolution $(1)\rightarrow(2)$, where the magnetic field is changed from $h_2$ to $h_1$. This evolution is characterized by $U= \mathcal{T}~{\rm Exp}\left( {-i\int^{h=h_1}_{h=h_2} H(t) dt}\right)$. Thus at $(2)$, before in contact to hot bath, density matrix $\rho^{(2)}=U \rho^{(1)} U^{\dagger}$, and energy $E^{(2)}= {\rm Tr}(\rho^{(2)} H^{(2)})$. Next, WM is made to equilibrate with hot bath of temperature $T_h=1/\beta_h$ and characterized by energy $E^{(3)}={\rm Tr}(\rho^{(3)}H^{(3)})$, where $\rho^{(3)}= e^{-\beta_h H^{(3)}}/{\rm Tr}(e^{-\beta_h H^{(3)}})$. The system is then adiabatically evolved to $(4)$ by $\tilde{U}= \mathcal{T}~{\rm Exp}\left({-i\int^{h=h_2}_{h=h_1} H(t) dt}\right)$, after detaching from the hot bath. Heat exchanged during hot and cold thermalization processes, $Q_h$ and $Q_c$ respectively, and efficiency $\eta$, are defined as,
\begin{eqnarray}\label{equation_qh_qc_eta}
	&&Q_h=E^{(3)}-E^{(2)},Q_c=E^{(1)}-E^{(4)}, \eta=\frac{(Q_h+Q_c)}{Q_h}.\qquad
\end{eqnarray}
The work done during the cycle is defined as $\omega=-(Q_h+Q_c)$, using the sign convention of $+(-)$ when energy is absorbed (dissipated) by the WM. Depending on the property of WM, following Clausius inequality \cite{single_qubit}, QOE can operate in four modes: (i) heat-engine [$\mathcal{E}$]: when $Q_h\geq0$, $Q_c\leq0$, $\omega\leq0$, (ii) refrigerator [$\mathcal{R}$]: when $Q_h\leq0$, $Q_c\geq0$, $\omega\geq0$, (iii) heat-accelerator [$\mathcal{A}$]: when $Q_h\geq0$, $Q_c\leq0$, $\omega\geq0$, and (iv) heater [$\mathcal{H}$]: when $Q_h\leq0$, $Q_c\leq0$, $\omega\geq0$. We note that these four operational modes can be explained in terms of energy diagram, and a fictitious effective temperature, which we define at any point on the Otto cycle as, ${\rm Tr} \left(\rho(t)H(t)\right)$ $=$ $E(t)$ $:=$ ${\rm Tr}\left(H(t)e^{-\beta_{\rm eff}H(t)}\right)/{\rm Tr}\left(e^{-\beta_{\rm eff}H(t)}\right)$. $T_{\rm eff}$ $=$ $1/\beta_{\rm eff}$ is a measure of how much an equilibrium temperature is required to equilibrate the Hamiltonian at time $t$ to have the same energy as we get at that particular point of the Otto cycle. This indirectly measures the effect of the driving field, which changes the system's internal energy. 
\\\\\indent

We use the full exact diagonalization method to obtain these clusters' energy spectra and eigenvectors. Using that, we calculate the density matrix, energy after each stroke, the amount of heat exchanged while attached to the thermal baths, and the work done on or by the system during two unitary processes. Following the Suzuki-Trotter decomposition, the unitary evolutions in two adiabatic paths ($(1)\rightarrow(2)$, and $(3)\rightarrow(4)$) are obtained by discretizing the full time scale in small pieces and assuming the Hamiltonian to be constant during these short time intervals.


\begin{figure}\centering
	\includegraphics[width=\columnwidth,height=!]{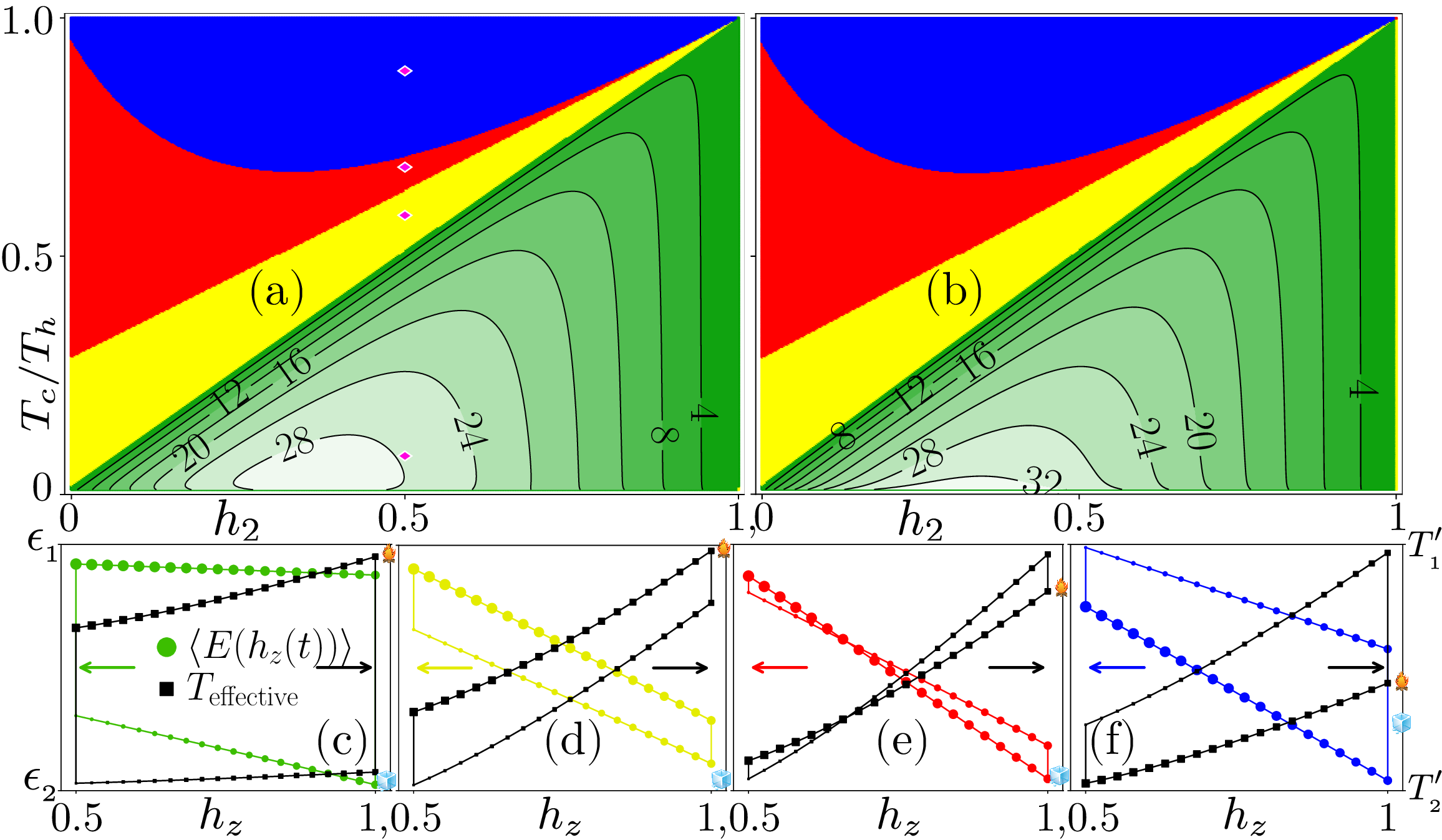}
	\caption{Operational regions for 8-site cluster in absence of Heisenberg interaction with (a) $\kappa=+1$, and (b) $\kappa=-1$, at $T_h=10^{2}$ and $h_1=1.0$ in both panels. Green, blue, yellow, and red region describes a heat engine, a refrigerator, a heat-accelerator, and a heater, respectively. For the heat-engine, efficiency is mentioned in contours (for better visibility, it is multiplied by 100). In panels (c-f), the colored plot shows the energy of the working medium along the path $1 \rightarrow 2 \rightarrow 3 \rightarrow 4$, the smallest circle being the starting point. The black contour shows the estimated effective temperature. (c-f) panels correspond to the specific $(T_c,h_2)$ values as indicated with magenta diamonds {\color{magenta}\rotatebox{90}{$\blacklozenge$}} in (a). For visual clarity, the vertical axes of (c-f) have been rescaled. Ticks of the energy axis correspond to $[\epsilon_{_1},\epsilon_{_2}]$ = (c) $[0,-0.11]$, (d) $[-0.008,-0.016]$, (e) $[-0.008,-0.014]$, (f) $[-0.007,-0.014]$. Ticks of the effective temperature axis correspond to $[T'_{_1},T'_{_2}]$ = (c) $[105,5]$, (d) $[101,57]$, (e) $[108,66]$, (f) $[140,69]$. On the $T_{\rm eff}$ axis, the hot and cold bath temperatures are indicated with fire \includegraphics[height=1em]{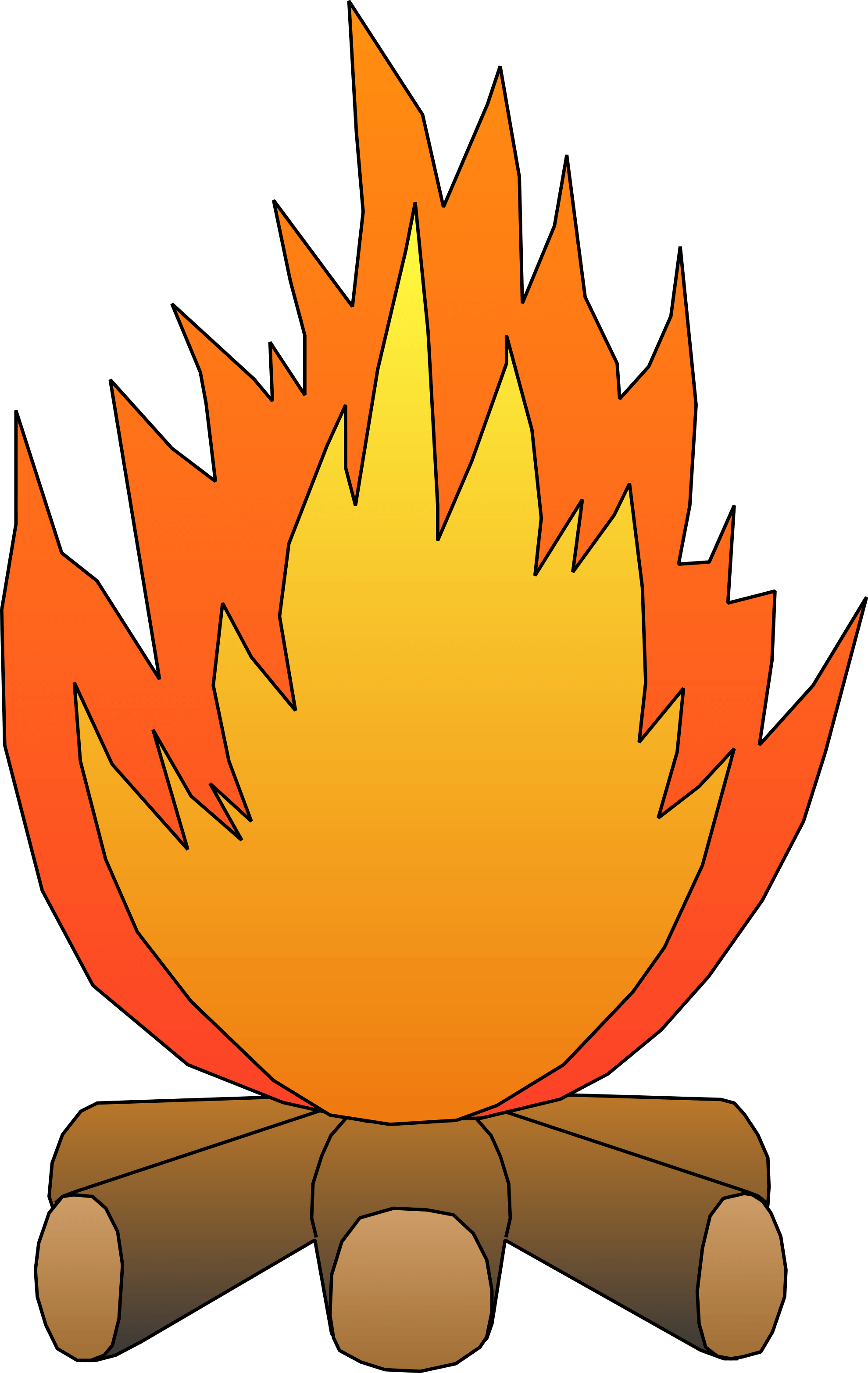} and ice \includegraphics[height=1em]{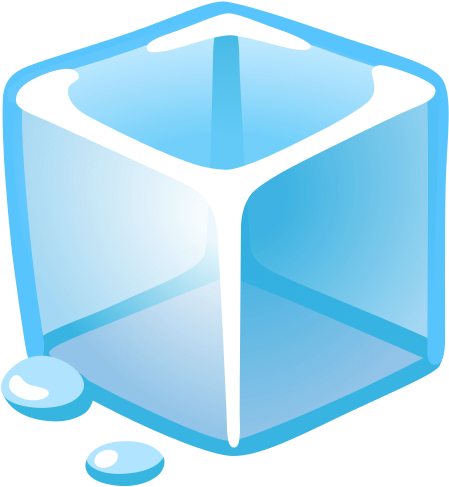} icon respectively.}\label{figure_working_modes}
\end{figure}

\section{Results}\label{section_results}

We chart out the different results in this section. Being qualitatively similar for all the clusters, we provide the results for the 8-site cluster only. The results tabulated here are obtained with $T_h=10^2$ and $h_1=1$. Considering $\kappa=5.5$ meV~\cite{abanerjee_2016}, these correspond to $T_h\sim63$ Kelvin, and $h_1\sim23$ Tesla.

\subsection{Working Regions}\label{subsection_working_modes}

In FIG.~\ref{figure_working_modes}, we show how the WM operates itself as $\mathcal{H,~R,~A,~E}$. The plots remain qualitatively unchanged in the presence of small $J$; only the area of these modes in the parameter space changes. The performance of QOE in these modes can be described by the average energy $\langle E\rangle$, or the effective temperature $T_{\rm eff}$ as introduced before. We choose one representative point from each of these four modes (magenta diamonds {\color{magenta}\rotatebox{90}{$\blacklozenge$}} in FIG.~\ref{figure_working_modes}a), and plot $\langle E\rangle$ and $T_{\rm eff}$ along the Otto cycle in FIG.~\ref{figure_working_modes}(c-f). At low $T_c$, WM experiences $\mathcal{E}$. When $T_c$ is increased, WM enters into $\mathcal{A}$ mode, which is characterized by $\omega > 0$ which implies that $|Q_c| > |Q_h|$, while $Q_h$ and $Q_c$ remain positive and negative respectively. This indicates that for a given $T_h$, as one increases $T_c$, the thermal fluctuation enhances $|Q_c|$. When $T_c$ is increased further, we enter into $\mathcal{H}$, where $Q_h$ becomes negative in contradiction with $\mathcal{E}$ and $\mathcal{A}$. It is caused by the effect of adiabatic pumping; the system's internal energy is changed to such an extent that the hot bath effectively becomes a sink. $T_c$ is further increased to reach $\mathcal{R}$, where $Q_c$ is positive, implying the WM is pumping out energy from the cold bath and transferring it to the hot one.
\\\\\indent
When we take sufficiently small $h_1$ to run the QOE deep inside the p-QSL region along the field axis~[FIG.~\ref{figure_bp},\ref{figure_mz}], vanishingly small efficiency and work are obtained, that too for high $T_h$ only; so this is outside p-QSL along the temperature axis. If we restrict ourselves only to p-QSL (very low $T_h$, small $h_1$), the WM does not work as $\mathcal{E}$ at all. All these indicate that there is no advantage in QOE due to the p-QSL. But this is even more interesting. That means a Kitaev candidate material, even if it fails to achieve the QSL due to the presence of other non-Kitaev interactions, can still be used as a QHE. In the following section, we show that the presence of some non-Kitaev (Heisenberg) interaction, which is unavoidable in real-world Kitaev materials, is not only acceptable but more desirable to build QOE. Interestingly, high temperature and large magnetic field are the suitable parameter space to achieve the QOE made with Kitaev WM, where $h_2\lesssim h_1$ creates the $\mathcal{E}$ mode, and $T_c\lesssim T_h$ generates the $\mathcal{R}$ region.

\begin{figure}\centering
	\includegraphics[width=\columnwidth,height=!]{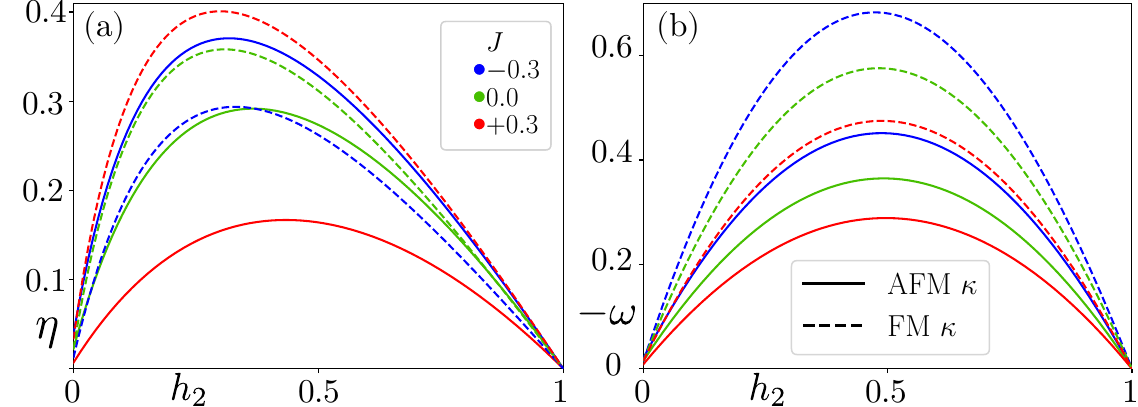}
	\caption{(a) Efficiency and (b) work, in the 8-site Kitaev cluster with Heisenberg interaction $J$ as in the legend box of (a). Solid and dotted lines correspond to the AFM and FM Kitaev model, respectively. The parameter values are $T_h=10^2,~T_c=10^0,~h_1=1$.}\label{figure_efficiency_and_work_for_Kitaev}
\end{figure}

\subsection{Effect of Heisenberg interaction}\label{subsection_heisenberg_interaction}
\begin{figure}\centering
	\includegraphics[width=\columnwidth]{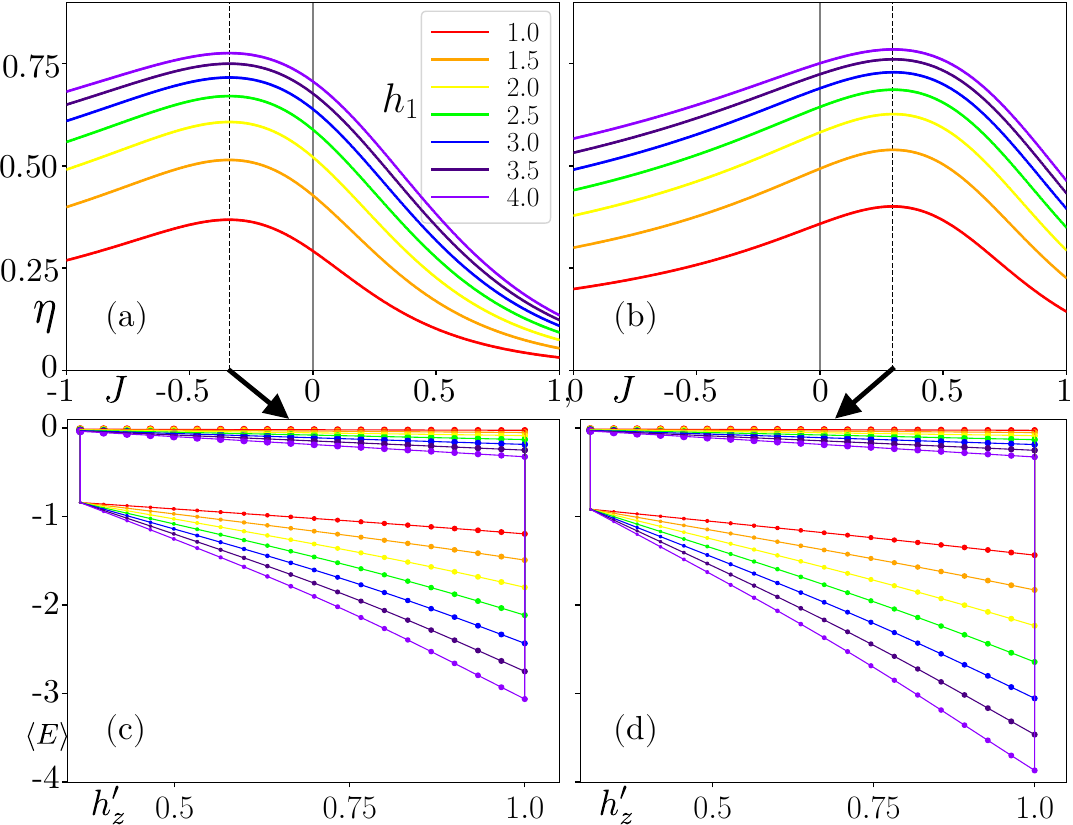}
	\caption{Efficiency in (a) AFM and (b) FM Kitaev model for varying $J$, with $h_2=0.365$ and $h_2=0.31$ respectively, for different $h_1$ as in the legend box. Note that the peak of $\eta$ remains at the same $J$ points (dotted black lines) as we increase $h_1$, and around these $J$ points, narrow band formation of the eigen-states occurs, as shown in FIG.~\ref{figure_eigenvalue_spectra}. In (c,d), we show the energy cycle at these maximum efficiency providing $J$ points for AFM and FM Kitaev model, respectively, where the $h_z$ axis is rescaled as $h_z'$ = $h_2$ + $(1-h_2)$ $\frac{(h_z-h_2)}{(h_1-h_2)}$ for better visual comparison. The point size increases as we walk along the steps 1 $\rightarrow$ 2 $\rightarrow$ 3 $\rightarrow$ 4 of FIG.~\ref{figure_otto_cycle}. Notice that for different $h_1$ values, WM radiates almost the same amount of heat to the cold bath, but the heat absorbed from the hot bath increases with an increase in $h_1$, enhancing the efficiency substantially.}\label{figure_heisenberg_efficiency}
\end{figure}

As this work is primarily focused on the $\mathcal{E}$ mode, we describe here the efficiency and work obtained in this mode, and present the effects of finite $J$. In FIG.~\ref{figure_efficiency_and_work_for_Kitaev}, we see that the pure FM Kitaev model has a better efficiency than the pure AFM one. However, in both cases, when we introduce $|J|=0.3$ with a sign opposite to that of $\kappa$, the efficiency increases. But, the work obtained is maximized (minimized) if the sign of $J$ is $-(+)$, irrespective of $\kappa$. In search of the reason behind efficiency enhancement for competing $\kappa$ and $J$, we pick $h_2$ values from FIG.~\ref{figure_efficiency_and_work_for_Kitaev}a, at which the pure Kitaev WM produces the highest efficiency, and plot its variation for a range of $J:-1\rightarrow+1$. The results are shown in FIG.~\ref{figure_heisenberg_efficiency}(a,b) with $\kappa=\pm1$ for different $h_1$ values.
\\\indent

It is remarkable that the plots are qualitatively symmetric when the signs of $(\kappa,J)$ change together. Efficiency, in absence of $J$, attains maximum at $h_2=0.365~{\rm and}~0.31$, for $\kappa=\pm1$. The obtained $\eta$ are even more enhanced for a competing $J$, and become maximum at $J=-0.34~{\rm and}~0.295$, for $\kappa=\pm1$, respectively. To understand the efficiency enhancement of KH system at some optimum combination of parameters, we compare the average energy calculated at any given point of the Otto cycle by plotting $\langle E(h_z(t))\rangle= {\rm Tr}(\rho(t) H(t))$ in FIG.~\ref{figure_heisenberg_efficiency}(c,d) for $\kappa=\pm1$, with respective $J$ at which maximum efficiency occurs. From Eq.(\ref{equation_qh_qc_eta}), the expression of efficiency reads as $\eta= 1+ (E^{(1)}-E^{(4)})/(E^{(3)}-E^{(2)})$, which indicates that the $\eta$ is mainly determined by $E^{(1)}$ and $E^{(2)}$, as $E^{(3)}$ and $E^{(4)}$ do not change much for different $J$. Near the optimum efficiency delivering $J$ points, $Q_h$ attains a very low value while delivering a similar amount of work; this makes the efficiency to be enhanced substantially. At this optimum $J$, as we increase $h_1$, the magnitude of $E^{(2)}$ increases while $E^{(3)}$ changes by a small amount, which directly enhances efficiency. The change in energy $E^{(1)}\rightarrow E^{(2)}$ and $E^{(3)}\rightarrow E^{(4)}$ actually denotes the energy released or absorbed by the system (in terms of work, to or from the external drive).
\begin{figure}\centering
	\includegraphics[width=\columnwidth,height=!]{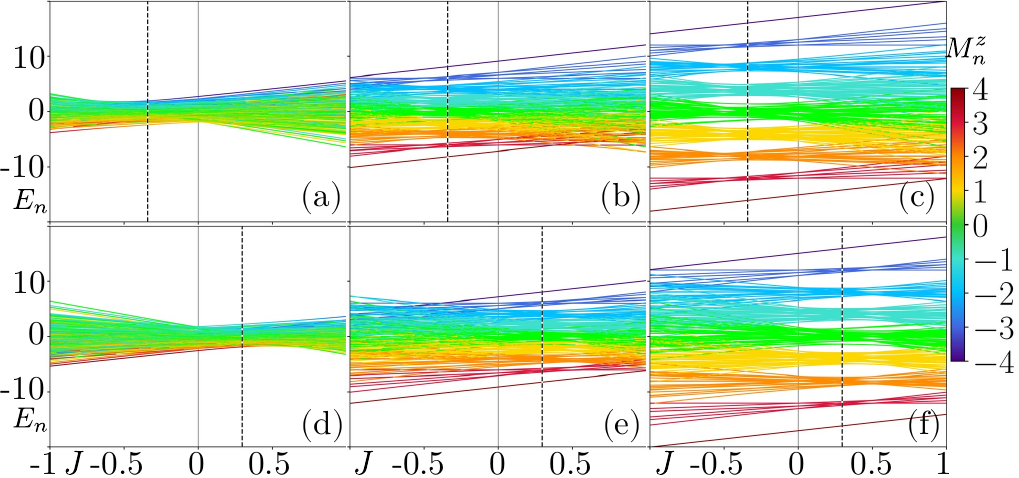}
	\caption{Energy eigen-spectra plots for (a-c) AFM and (d-f) FM Kitaev model against $J$. The $h_z$ values are $0.365$ at (a), $0.31$ at (d), $2.0$ at (b,e), and $4.0$ at (c,f). Black dotted line is at (a-c) $J=-0.34$, and (d-f) $J=+0.295$, corresponding to maximum efficiency delivering $J$ points in AFM and FM Kitaev model respectively. Multiple narrow bands are formed around these dotted lines as one increases $h_z$. The color code in an eigen-energy denotes the total magnetization in $\hat{z}$-direction for that state. The number of bands is equal to the possible values of total magnetization, which turns out to be constant for a given band.}\label{figure_eigenvalue_spectra}
\end{figure}
\\\\\indent

\begin{figure}\centering
	\includegraphics[width=\columnwidth,height=!]{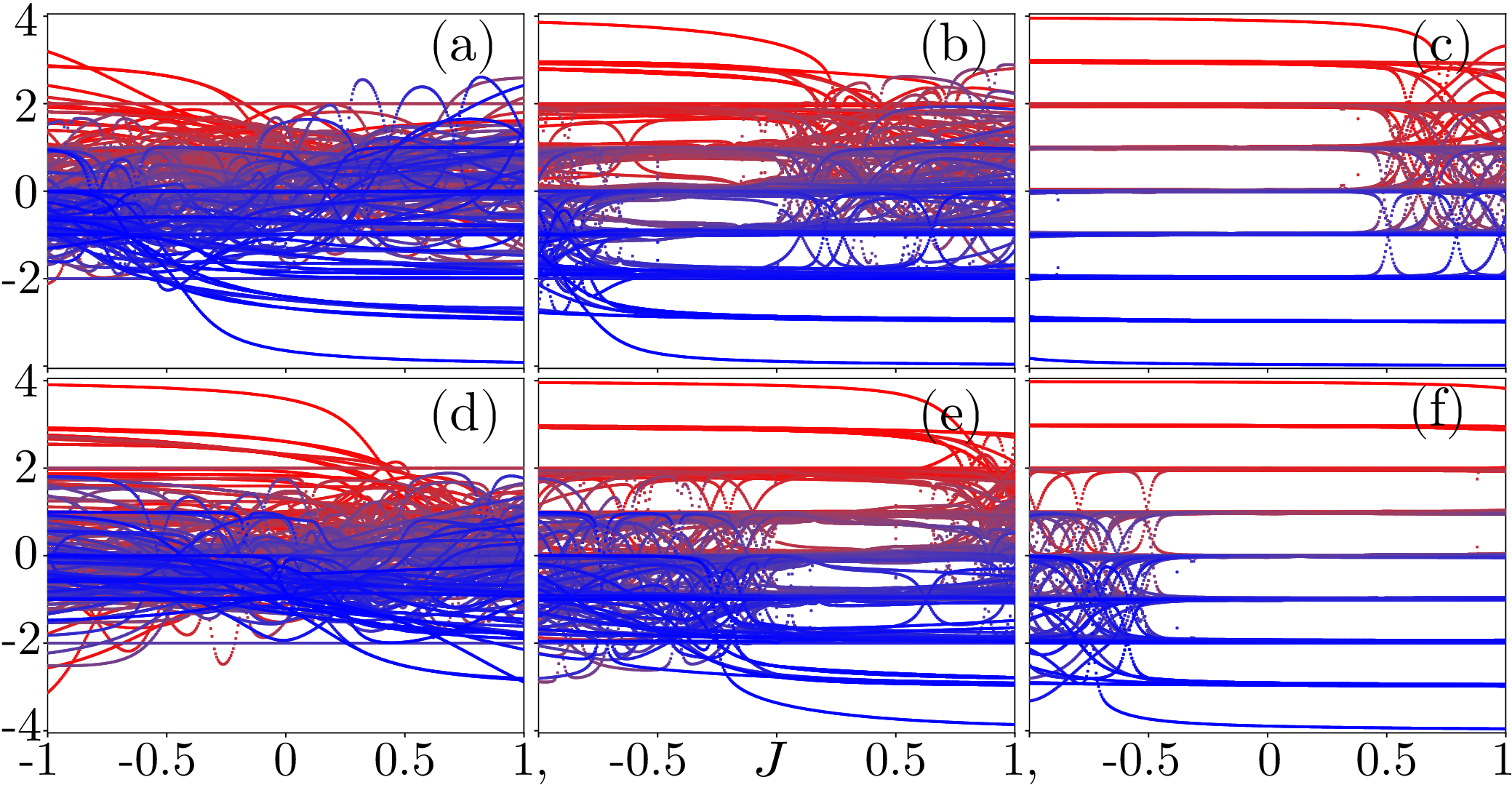}
	\caption{Magnetization at each of the eigen-states for (a-c) AFM and (d-f) FM Kitaev model against $J$. The $h_z$ values are $0.365$ at (a), $0.31$ at (d), $1.0$ at (b,e), and $2.0$ at (c,f). Pure red ({\color{red}$\blacksquare$}) and pure blue ({\color{blue}$\blacksquare$}) represent the lowest and highest eigenstates' magnetization, and the color interpolates between these two for intermediate states.}\label{magnetization_point_plot}
\end{figure}
{\it Realization of narrow bands and emergent magnons:} Our numerical calculation reveals that for certain combinations of $\kappa, J, h_z$, the complete set of eigenstates along the unitary stroke remains quasi-stationary. Concomitantly, the eigenstates form multiple narrow bands as shown in [FIG.~\ref{figure_eigenvalue_spectra}] and the number of bands equals the possible quantized values that the total $S^z$ can take. This implies that for the 8-spin cluster, there should be nine bands, and our numerics exactly realize that. We infer this fact by investigating the eigenvalue spectrum and the expectation value of total magnetization denoted by $M_n^z$ $:=\expval{\hat{M}_z}{n}$ = $\expval{\sum_{j=1}^{N}S_j^z}{n}$, calculated at $\ket{n}^{\rm th}$ eigen-state. We see a gradual transition of the spectrum from continuous to a closely packed multiple narrow bands as one increases the value of the external magnetic field. The expectation values of $M_z$ gradually evolve from arbitrary values to quantized values. In [FIG.~\ref{magnetization_point_plot}], we plot the $M^z$ values obtained, and it clearly demonstrates that in the region where narrow band formation occurs, magnetization values become quantized. It is known that when the magnetization is quantized exactly to the expected total $S^z$, it can be well described by magnons. Thus, our results point out another scenario where the emergent magnons play a crucial roles in maximizing the efficiency and offer another avenue of efficiency increasing mechanism in addition to adiabaticity~\cite{victor_2020_adibaticity,quantum_lubrication}, or long range interactions~\cite{long_range_advantage}. Interestingly, we observe that the system exchanges more energy for such a scenario as established in panels (c) and (d) in FIG.~\ref{figure_heisenberg_efficiency}. 
\begin{figure}\centering
	\includegraphics[width=\columnwidth,height=!]{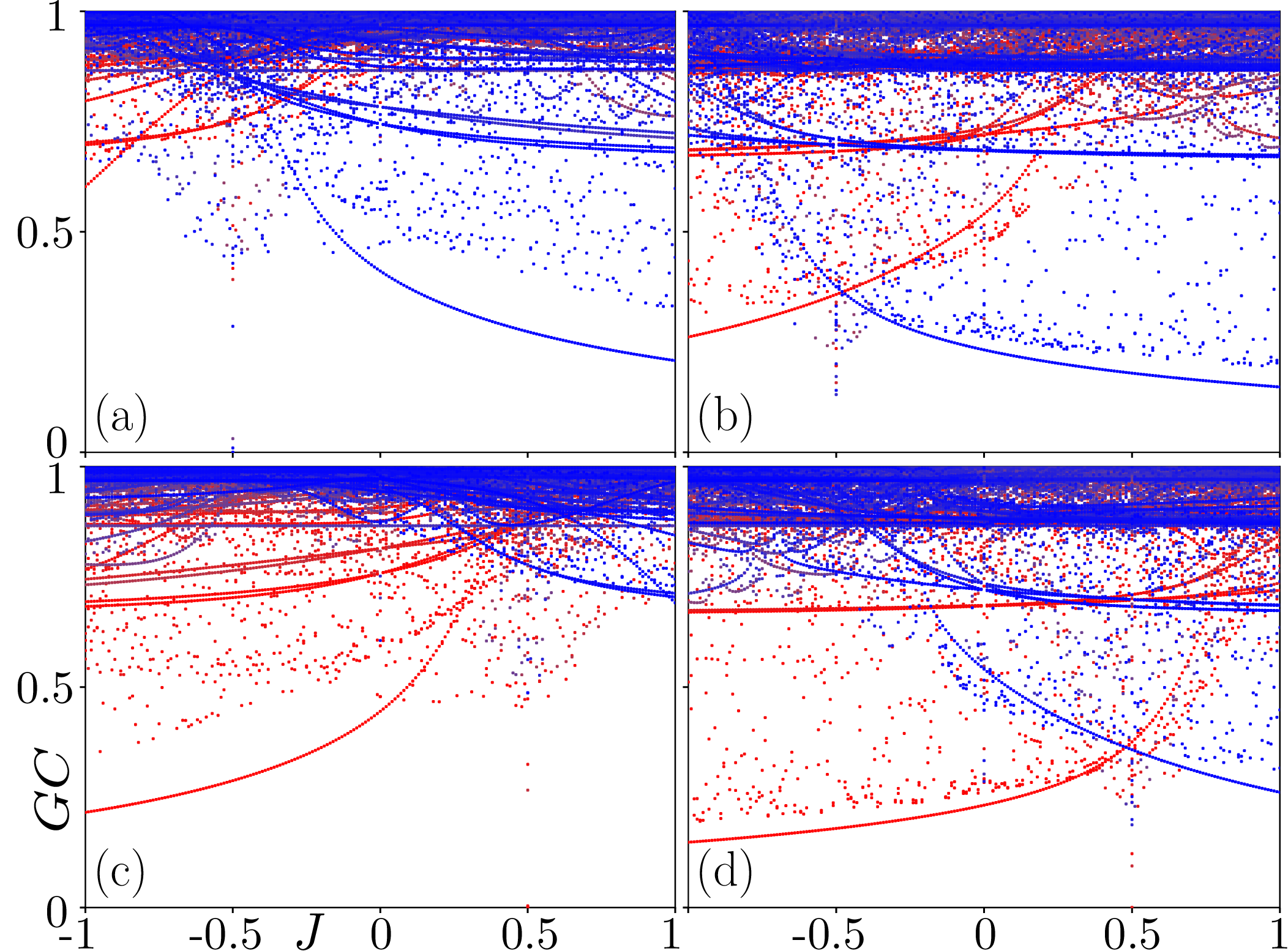}
	\caption{The generalized concurrence (GC) for all eigen-states of the 8-site cluster is plotted. Pure red ({\color{red}$\blacksquare$}) and pure blue ({\color{blue}$\blacksquare$}) represent the lowest and highest eigen-state concurrence, and the color interpolates between these two for intermediate states. The upper and lower panels represent AFM and FM $\kappa$, respectively. $h_z$ is taken $0.365$ at (a), $0.31$ at (c), and $1.0$ at (b,d). }\label{concurrence}
\end{figure}
\\\\\indent
{\it Possible role of entanglement:} We end this discussion with a possible role of entanglement contributing to observed results. The paths $(1)\rightarrow(2)$ and $(3)\rightarrow(4)$ of the Otto cycle are unitary, and the entanglement entropy does not change along these paths. There is a discontinuous jump in entropy from $(2)\rightarrow(3)$ and $(4)\rightarrow(1)$. It is interesting to note that entanglement entropy at point $(1)$ (and $(2)$) follows a similar pattern as $E^{(1)}$ (and $E^{(2)}$) while varying $J$ for fixed $\kappa,~h_2,~h_1$. Entropy at points $(3)$ and $(4)$ also follows a similar pattern as $E^{(3)}$ and $E^{(4)}$, but the magnitude is reduced due to the high temperature of the hot bath. Another fascinating observation is that the generalized concurrence~\cite{Dutta2019} (a measure of entanglement for pure states) is symmetric [FIG.~\ref{concurrence}] in the KH system for similar combinations of $(\pm\kappa, \mp J)$, similar to what we observe in FIG.~\ref{figure_heisenberg_efficiency}. Notice in FIG.~\ref{concurrence}(b,d), ground state concurrence is very close to that of the highest excited state, when we flip the sign of $\kappa$ and $J$ both. The QOE results are not exactly the same when we flip $(\pm\kappa, \mp J)$, because the response due to AFM and FM couplings is different in the presence of a magnetic field. QOE uses the full energy spectra to determine its performance, and it is the concurrence of all states that is important. The entanglement results indicate a possible duality in the KH system for competing $\kappa$ and $J$ in the presence of a magnetic field, which shows its effect in determining the efficiency of QOE made with this WM.

\subsection{Large-$S$ heat-Engine}\label{subsection_large_s}
In recent years, many theoretical and experimental studies have been conducted for large spin-$S$ Kitaev and KH models~\cite{PhysRevLett.130.156701, PhysRevLett.123.037203, PhysRevLett.124.087205, PhysRevB.107.014411, PhysRevB.78.115116, PhysRevB.98.214404,doi:10.7566/JPSJ.87.063703, PhysRevResearch.2.022047}.
Theoretically, it offers a unique platform to study how strong quantum fluctuations realized for the QSL evolve as the magnitude of spin itself becomes larger~\cite{PhysRevB.102.155134}. This gives a direct way of measuring how a larger magnitude of $S$, known for weaker quantum fluctuation, competes with frustrations induced by interactions. Experimentally, there are also materials which propose the KH model with higher $S$~\cite{PhysRevLett.124.087205}. There have been very few studies on QOE for either large spin or for multilevel systems with different dependency of efficiency on the magnitude of spin or the number of levels~\cite{Hartmann2020multispincounter,PRXQuantum.3.040334,Bouton2021QuantumHeatEngine,e24020268,PhysRevE.92.022142,PhysRevE.102.022143}. Here we briefly present the large-$S$ QOE performance for the Kitaev model Hamiltonian considered in Eq.(\ref{hamiltonian}). The algebra of different components of large spin $\vec{S}$($\{S_j^x,S_j^y,S_j^z\}$) now obey the commutations $[S_j^\alpha,S_j^\beta]=i\epsilon^{\alpha\beta\gamma}S_j^\gamma$ with a Hilbert space of dimension $(2S+1)$ per spin~\cite{Jahromi2024}. As the exact diagonalization technique has been used, we restrict ourselves to $N=4$ only.
\begin{figure}\centering
	\includegraphics[width=\columnwidth,height=!]{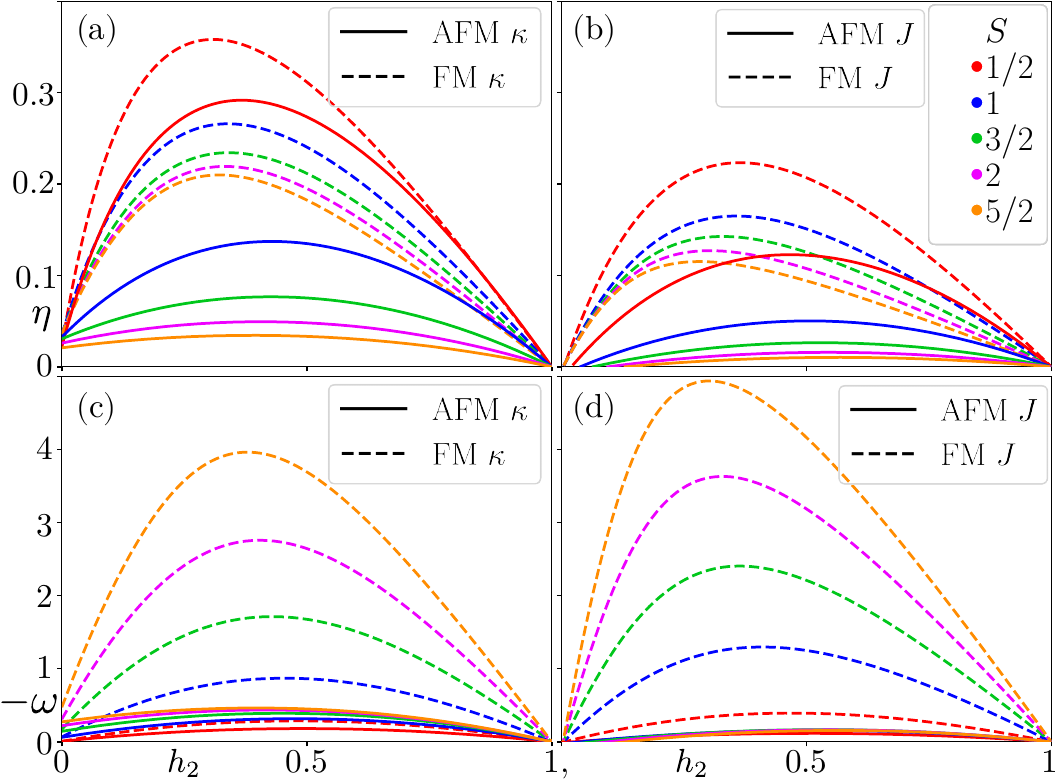}
	\caption{(a,b) Efficiency $\eta$, and, (c,d) work ($-\omega$) delivered by a spin-$S$ heat-engine for the 4-site cluster. We have taken the WM to be a (a,c) pure Kitaev cluster with $|\kappa|=1$, and (b,d) pure Heisenberg (tri-coordinated) cluster with $|J|=1$. The parameter values are set at $(T_h,T_c,h_1)=(10^{2},10^{0},1)$, and $h_2$ varies from $0\rightarrow h_1$. In all of the panels, different colors are for different spin sizes, as indicated in the legend box of (b).} \label{figure_large_S_efficiency}
\end{figure}
\\\\\indent
In FIG.~\ref{figure_large_S_efficiency}, we plot the efficiency and work output for different magnitudes of $S$ (up to $S$=5/2). We observe that the FM Kitaev model yields better efficiency than the AFM one, for a given $S$. However, in both cases, the efficiency decreases with increasing $S$, thus showing evidence of quantum advantage. The work output for the FM Kitaev model is higher than the AFM one, and in both cases, it increases with $S$. To get an asymptotic dependency of efficiency and work, we fit the maximum of $\eta$ and $\omega$, with orders of $S^{-1}$. For AFM and FM Kitaev model, the maximum efficiency fits as $\eta_{\rm max}= 0.1615 S^{-1} - 0.0295$, and $\eta_{\max}= 0.0928 S^{-1} + 0.1729$, respectively. The maximum efficiency delivering points do not coincide with the maximum work producing $h_2$ values. The maximally produced work depends on $S$ as, $-w_{\rm max}=0.49 (1- \rm exp(-1.11(S-0.09)))$ for AFM Kitaev model, and $-w_{\rm max}=-0.13 + 0.61S + 0.41 S^{2}$ for the FM one. It is interesting that while an AFM Kitaev model gives a saturated value of work on increasing $S$, the FM Kitaev model yields a polynomial dependency.
\\\\\indent
In FIG.~\ref{figure_large_S_efficiency}(c,d), we plot the efficiency and work of a pure Heisenberg model on the 4-site cluster. We observe that the same quantum advantage persists in efficiency, and there is a dimensional advantage in work. Interestingly, the Kitaev WM is much more efficient than the Heisenberg cluster, at the expense of a lesser disadvantage in work. This indicates the promising usefulness of the frustrated Kitaev system as compared to the Heisenberg model, in a quantum engine.


\section{Conclusion}\label{section_conclusion}
We have investigated a quantum Otto engine with Kitaev-Heisenberg clusters serving as the many-body working medium. A time-dependent Zeeman field along the $\hat{z}$ direction is taken as an external drive. A pure AFM or FM Kitaev model shows all four different modes the system can act as: heat-engine, refrigerator, heat-accelerator, and heater. Interestingly, the regions are almost cluster-size independent, which indicates the possible realization of similar results in larger Kitaev clusters too, which are computationally expensive. Heisenberg interaction added to either FM or AKM Kitaev model destroys the proximate-QSL regime to a varying degree, but it increases the efficiency only when the sign of a small $J$ is opposite to that of $\kappa$. It possibly indicates that the proximate-QSL property is not the sole efficiency-determining factor. Neither does it depend on the simple quantum ordering mechanism between the working media and the external drive, as evident from the fact that for $\kappa$ negative and $J$ positive yields a better efficiency in comparison to $J$ negative. Rather, it is the mutual quantum coherence between the external drive and the relative contribution from frustration and the collinear ordering that optimizes the efficiency. Our work calls for a more in-depth analysis of this mechanism.
\\\\\indent
The present study uncovers a very subtle many-body effect of the Kitaev-Heisenberg system under a Zeeman field. The eigen-spectrum forms dense and narrow bands for a certain combination of signs and magnitude of $J, ~\kappa$, and $h_z$. Interestingly, when the working media dynamically enters these combinations of parameters, it realizes maximum efficiency. This connects the equilibrium property of the system to be a key efficiency-determining factor. The number of bands is equal to the quantized value of $S^{z}_{\rm total}$, and the value is fixed at a particular band. The quantization of $S^z_{\rm total}$ indicates that the system can be effectively described by magnons, and it offers an interesting mechanism of weakly interacting magnons as an efficiency increasing mechanism. Further, we were able to characterize the various working modes by introducing the concept of fictitious effective temperature, which accounts for the energy exchange due to the external drive. Interestingly, this effective temperature also correctly captures the maximum efficiency where energy loss due to the external drive is mimimized. Lastly, we have studied the heat-engine for a spin-$S$ Kitaev cluster up to $S=5/2$, and found a quantum advantage in efficiency, with $1/S$ dependency for both AFM and FM Kitaev models. The maximum work output exponentially saturates with increasing $S$ for the AFM Kitaev model, whereas, it has a polynomial of second order dependency for the FM one. We expect that our study sheds new understanding on how the efficiency of a quantum engine depends crucially on multiple mechanisms such as competing interactions, frustrations, and duality of eigen-states, as well as emergent weakly interacting magnons. All of these facts act crucially to determine the engine performance for a many-spin system.
\\\\\indent
Given the recent interesting findings on the Kitaev-Heisenberg model~\cite{PhysRevB.109.L220403}, or Kitaev model under external Zeeman field~\cite{PhysRevB.108.165118}, and other magnetic models with competing ground states~\cite{PhysRevB.102.224404,PhysRevB.107.134438}, our study calls for quantum engine studies in these systems to explore the effect of entanglement~\cite{PhysRevB.94.045421}, dimensions~\cite{naveen_2008} and other aspects. Further, it remains to see whether the results obtained for small Kitaev-Heisenberg clusters are also valid in the thermodynamic limit and the exact quantum factor optimizing the efficiency of such engines. Interestingly, in thermodynamic limit, the Kitaev model defined on higher dimensions~\cite{naveen_2008}, or in different lattices with gapless Fermi surface or line~\cite{PhysRevResearch.3.L012001,Kells_2011}, may be of interest to see dynamical effect of defect density~\cite{PhysRevLett.100.077204,PhysRevB.78.045101,PhysRevB.102.134309} on quantum Otto engine and we leave it as future scope of study.


\section*{Data Availability Statement}
The data supporting this study's findings are available upon reasonable request from the corresponding author.


\section*{Acknowledgement}
The authors acknowledge valuable discussions with Victor Mukherjee. S.M.P. also thanks Soumya Bera, Debashish Mondal, and Ramita Sarkar for interesting discussions. Numerical calculations are partly performed in SAMKHYA (High-Performance Computing facility provided by the Institute of Physics, Bhubaneswar).

\bibliography{bibfile.bib}
\end{document}